\documentclass[amsfonts, twocolumn, prx, superscriptaddress,citeautoscript]{revtex4-1}
\usepackage{graphicx}
\usepackage{mathrsfs}
\usepackage{amsmath}
\usepackage{amssymb}
\usepackage{color}
\usepackage[nice]{nicefrac}
\usepackage{float}
\usepackage{dsfont}
\usepackage{lipsum}
\usepackage[a4paper,lmargin=2cm,rmargin=1.2cm,tmargin=2.2cm,bmargin=2.2cm]{geometry}

\usepackage{color}
\definecolor{darkred}{rgb}{0.6,0.,0.}
\definecolor{darkgreen}{rgb}{0.,0.5,0.}
\definecolor{darkblue}{rgb}{0.,0.,0.6}

\usepackage[
breaklinks,
colorlinks=true,
linkcolor=darkred,
citecolor=darkgreen,
urlcolor=darkblue]{hyperref}

\begin{document}

\def\be{\begin{eqnarray}}
\def\ee{\end{eqnarray}}
\def\nn{\nonumber}

\title{Nonstoquastic Hamiltonians and quantum annealing of an Ising spin glass}
\author{Layla Hormozi}
\affiliation{Center for Theoretical Physics and Research Laboratory of Electronics, Massachusetts Institute of Technology, 77 Massachusetts Avenue, Cambridge, MA 02139, USA}
\author{Ethan W. Brown}
\affiliation{Theoretical Physics and Station Q Zurich, ETH Zurich, 8093 Zurich, Switzerland}
\affiliation{Mindi Technologies Ltd., 71-75 Shelton Street, Covent Garden, London, WC2H 9JQ, United Kingdom}
\author{Giuseppe Carleo}
\affiliation{Theoretical Physics and Station Q Zurich, ETH Zurich, 8093 Zurich, Switzerland}
\author{Matthias Troyer}
\affiliation{Theoretical Physics and Station Q Zurich, ETH Zurich, 8093 Zurich, Switzerland}
\affiliation{Quantum Architectures and Computation Group, Microsoft Research, Redmond, WA 98052, USA}

\date{\today}

\begin{abstract}
We study the role of Hamiltonian complexity in the performance of quantum annealers. We consider two
general classes of annealing Hamiltonians: stoquastic ones, which can be simulated efficiently using the quantum
Monte Carlo algorithm, and nonstoquastic ones, which cannot be treated efficiently. We implement the latter
by adding antiferromagnetically coupled two-spin driver terms to the traditionally studied transverse-field Ising
model, and compare their performance to that of similar stoquastic Hamiltonians with ferromagnetically coupled
additional terms. We focus on a model of long-range Ising spin glass as our problem Hamiltonian and carry out
the comparison between the annealers by numerically calculating their success probabilities in solving random
instances of the problem Hamiltonian in systems of up to 17 spins. We find that, for a small percentage of
mostly harder instances, nonstoquastic Hamiltonians greatly outperform their stoquastic counterparts and their
superiority persists as the system size grows. We conjecture that the observed improved performance is closely
related to the frustrated nature of nonstoquastic Hamiltonians
\end{abstract}

\maketitle

\section{Introduction}

Physically-inspired approaches play a prominent role in both analyzing and devising solution strategies to complex optimization problems. For example, a large number of combinatorial optimization problems can be encoded into the couplings of Ising Hamiltonians, such that the minimum-energy configuration of the latter corresponds to the optimal solution of the former~\cite{fu86, mezard87, mezard09, lucas14}. In principle, at low enough temperatures these physical systems should eventually relax to their ground state, which subsequently can be measured and decoded to provide a solution to the original optimization problem. In reality, however, the relaxation time can be extremely long. In the language of disordered Ising models, the hardness of the encoded optimization problems can be attributed to the rough shape of the energy landscapes of the corresponding Hamiltonians in the configuration space, which typically consist of many hills and valleys~\cite{mezard09}. The presence of these local extrema renders the task of finding the global minimum of the system (i.e. the true ground state) very difficult. 

To overcome this problem, quantum annealing was first introduced as a computational simulation method, similar to simulated annealing~\cite{SA}, but with quantum fluctuations taking the place of thermal fluctuations~\cite{nishimori98}. The idea of quantum annealing is then to use quantum fluctuations to allow the system to tunnel through `spiky' barriers, for which simulated annealing is inefficient, thereby improving the system's chance to explore the configuration space more efficiently. Similar to simulated annealing, in this case the strength of the fluctuations is gradually reduced to zero, allowing the system to relax into the ground state of the problem Hamiltonian.  

A quantum annealing device is a machine that physically implements this approach by realizing a time-dependent Hamiltonian, which attempts to follow the adiabatic quantum algorithm~\cite{farhi00, farhi01, das08, fn:zeroT}. This machine is initialized in the ground state of a beginning Hamiltonian, then evolves in time while following the adiabatic path as closely as possible, to finally relax into the ground state of the problem Hamiltonian. The final ground state configuration can be subsequently measured to provide a solution to the encoded optimization problem. Following the recent technological advances in manufacturing systems of coupled qubits, the idea of building a special-purpose quantum annealing device to solve optimization problems has attracted much attention and prototypes of such devices have already been implemented~\cite{johnson11, boixo13, boixo14}. 

Recent studies of the performance of these quantum annealers, compared to quantum Monte Carlo (QMC) simulations, have shown that for tunneling between the local minima in the energy landscape, quantum annealing and QMC exhibit the same scaling of computational time with system size~\cite{isakov15, denchev15,  jiang16}. This observation has led to the conjecture that if QMC is inefficient in simulating a problem, then a quantum annealer is also inefficient in solving that problem so long as its Hamiltonian along the annealing path belongs to the class of the so-called stoquastic Hamiltonians, for which sign-problem-free QMC simulations can be performed. This conjecture implies that for a physical quantum annealing device to have any chance of out-performing classical algorithms (such as QMC), it must take advantage of nonstoquastic Hamiltonians, for which efficient QMC cannot be performed~\cite{bravyi06, loh90}.  

The formal definition of stoquastic Hamiltonians states that their path-integral configurations (in some local computational basis), contributing to the partition function, all have real and non-negative weights. For this to be true, it suffices to have matrix representations in the computational basis with real and non-positive off-diagonal matrix elements~\cite{bravyi06}. These Hamiltonians include bosonic problems, non-frustrated quantum magnets and certain special fermionic problems~\cite{loh90}. In general, for these systems QMC algorithms can efficiently update the path-integral configurations and propose new configurations with effort that only grows polynomially with the problem size~\cite{fn:qmcmarkov}. 

Path-integral QMC methods map the $d$-dimensional quantum system to a $(d+1)$-dimensional classical one. The quantum partition function can then be mapped to the partition function of $p$ copies of a classical system, which occupy an extra dimension, thus taking the form,
\be
Z = \text{tr} \: e^{-\beta H} \simeq  \text{tr} \: (e^{-\beta H/p})^p, 
\ee
where $\beta$ is proportional to the inverse temperature. This additional dimension can be interpreted as imaginary time with each time slice defined as,
\be 
\Delta\tau = \beta/p.
\ee
The partition function is then reduced to $p$ sums over complete sets of basis states, $\{l_1\}, ... \{l_p\}$, which are weighted by the size of the time slice and the off-diagonal matrix elements of $H$, 
\be
Z \simeq \prod_{j = 1}^p \sum_{l_j} \langle l_j |e^{-\Delta\tau H_{j, {j+1}}}| l_{j+1} \rangle.
\label{qmc}
\ee

When the off-diagonal matrix elements, $H_{j, j+1}$, are zero or negative, these weights are purely positive for each time slice, which in turn enables the stochastic sampling of these configurations in a QMC simulation. These Hamiltonians are dubbed `stoquastic'~\cite{bravyi06}, which combines the words `quantum' and `stochastic.' Here, for all practical purposes, the term `stoquastic' simply means `avoiding the sign problem.'~\cite{bravyi14-2}

For Hamiltonians whose matrix representations in the computational basis have positive or complex off-diagonal elements, the corresponding weights in Eq.~(\ref{qmc}) will be non-positive. These Hamiltonians are generally more complex than stoquastic ones~\cite{troyer05}, and they constitute an essential ingredient for universal adiabatic quantum computing~\cite{aharonov04, biamonte08}. 

Here we examine the potential power of this complexity in a different context and ask whether quantum annealers with nonstoquastic Hamiltonians can show superior performance as optimization machines. Along these lines, Ref.~\onlinecite{nishimori17} provides encouraging evidence that, for certain problems, nonstoquastic Hamiltonians can provide a scaling advantage over the traditionally-studied transverse-field annealing Hamiltonians.

To realize a concrete analysis, we pick a long-range Ising spin glass model as our problem Hamiltonian, choose a specific annealing schedule, fix a total annealing time and measure the performance of our nonstoquastic Hamiltonian by calculating success probabilities in a range of system sizes. 
In what follows, we first set the stage in Section~\ref{sec:notation} by briefly explaining the notation and the methods used. We then present the numerical results in Section~\ref{sec:results}, and conclude by presenting a discussion of our observations in Section~\ref{sec:discussion}. 

\section{Setting up the problem}
\label{sec:notation}

\subsection{The Notation}
The problem Hamiltonian, encoded as an Ising model, generally has the form,
\be
H_P = \sum_{i<j = 1}^N J_{ij} \sigma^z_i \sigma^z_j + \sum_{i = 1}^N h_i \sigma^z_i,
\ee
where the choices of pairwise couplings, $J_{ij}$, and the individually applied fields, $h_i$, determine the specific optimization problem of interest. Here we focus on a disordered spin glass problem that resembles the Sherrington-Kirkpatrick (SK) model~\cite{SK}. This model is infinite-dimensional in the thermodynamic limit and it has been shown that its worst cases are nondeterministic polynomially (NP)  hard~\cite{barahona82}. The problem is defined on a fully-connected graph, i.e. every pair of spins is coupled, and the parameters $h_i$ and $J_{ij}$ are randomly chosen from a continuous Gaussian distribution with zero mean and unit variance~\cite{fn:SK}. 

The original time-dependent Hamiltonian for the adiabatic quantum algorithm has the form~\cite{farhi00}, 
\be
H^0(\tau) &=& (1-\tau) H_B + \tau H_P,
\label{eq:HX}
\ee
where $\tau = t/T \in [0,1]$ is the dimensionless annealing parameter and $T$ is the total annealing time. The beginning Hamiltonian (at $t=0$), whose ground state is unique and easy to implement, is traditionally chosen to be the uniform transverse-field Hamiltonian,
\be
H_B =  \sum_{i = 1}^N  \sigma^x_i.
\ee
Each term in $H_B$ effectively flips a spin in the computational basis, thus it is a driver for quantum fluctuations, which allow the system to explore the energy landscape of the problem Hamiltonian during the annealing process. As $t$ increases, the strength of the driver terms decrease while the strength of the problem Hamiltonian increases. If this process is done slowly enough so that the adiabatic theorem can be applied, then at $t = T$ the ground state of the system should evolve into the ground state of $H_p$~\cite{farhi00}. 

The Hamiltonian $H^0$ (Eq.~(\ref{eq:HX})) is stoquastic but, by suitably modifying the driver terms, a nonstoquastic Hamiltonian can be obtained. In this work we use driver Hamiltonians that include terms of the form $\sigma^x_i\sigma^x_j$ with both antiferromagnetic and ferromagnetic couplings~\cite{biamonte08, seki12, Seoane12, crosson14, seki15}. To avoid the degeneracy resulting from the frustrated state of antiferromagnetically-coupled spins on a fully-connected graph, following Ref.~\onlinecite{farhi02}, we choose our total annealing Hamiltonian to be of the form, 
\be
H(\tau) = (1-\tau) H_B + \lambda \tau(1-\tau) H_I + \tau H_P,
\label{eq:genform}
\ee
where we begin with the unique ground state of $H_B$ and enter the additional coupled driver terms, $\sigma^x_i\sigma^x_j$, via the intermediate term, $H_I$. The parameter $\lambda$ can in general control the strength of $H_I$ but here is set to be $\lambda = 1$~\cite{fn:scaling}. 

Note that in the computational basis, the local effect of each $\sigma^x_i\sigma^x_j$ term is to flip the $i^{th}$ and $j^{th}$ spins simultaneously. To distinguish the effect of flipping a pair of spins (as opposed to a single spin flip due to $H_B$) from the possible specific effects of nonstoquasticity, we also consider an intermediate Hamiltonian with uniform ferromagnetic couplings. Thus, we end up with the following three intermediate Hamiltonians:
\be
H_I^F &=& - \sum_{i<j = 1}^N \sigma^x_i\sigma^x_j\label{eq:HiXX}\\
H_I^A &=& + \sum_{i<j = 1}^N \sigma^x_i\sigma^x_j,\label{eq:HiXXp}\\
H_I^M &=&  \sum_{i<j = 1}^N r_{ij} \: \sigma^x_i\sigma^x_j\label{eq:HiXXS}.
\ee
In the latter case $r_{ij}\in\{-1,1\}$ is randomly chosen, giving rise to an intermediate Hamiltonian with both ferromagnetic and antiferromagnetic couplings~\cite{fn:other}. Here the superscripts $F$, $A$ and $M$ refer to \emph{Ferromagnetic}, \emph{Antiferromagnetic} and \emph{Mixed-signed}, respectively.

Inserting either $H_I^A$ or $H_I^M$ in Eq.~(\ref{eq:genform}) results in nonstoquastic total Hamiltonians (for $\tau \neq 0, 1$), while inserting $H_I^F$ in Eq.~(\ref{eq:genform}) produces a stoquastic Hamiltonian with coupled spin flip driver terms. In what follows we will refer to the intermediate Hamiltonians as drivers with coupled fluctuations, or simply as coupled drivers. 

We compare the success rate of Hamiltonians with coupled drivers against that of the original Hamiltonian (Eq.~(\ref{eq:HX})), as our reference. To simplify referencing, we label the Hamiltonians with coupled drivers as,
\be
H^\alpha(\tau) &=& H^0(\tau) + \tau (1-\tau) H_I^\alpha,
\label{eq:HXX}
\ee
where $\alpha$ = $F$, $A$, $M$, correspond to stoquastic, nonstoquastic with uniform antiferromagnetic driver terms and nonstoquastic with mixed driver terms, respectively. In what follows we use the same index $\alpha$ for labeling various quantities such as success probabilities, $P^\alpha$, and minimum gaps, $\Delta^\alpha$, which result from $H^\alpha$. 

\subsection{Methods and Metrics}

Our main numerical tools are exact diagonalization, to calculate the instantaneous energy spectra of the Hamiltonians $H^0$ and $H^\alpha$, and the numerical solution of the time-dependent Schr{\"o}dinger equation, 
\be
\frac{i}{T}\frac{\partial} {\partial \tau}|\psi(\tau)\rangle = H(\tau) |\psi(\tau)\rangle,
\ee
to simulate the process of quantum annealing. Note that we have set $\hbar = 1$. As our main metric of performance we choose the success probability~\cite{crosson14}, which is defined as the square of the overlap between $|\psi_g\rangle$, the true ground state of $H_P$, obtained from exact diagonalization and $|\psi^\alpha_g(\tau = 1)\rangle$, the approximate ground state of $H_P$, resulting from numerically solving the time-dependent Schr{\"o}dinger equation associated with $H^\alpha$, i.e., 
\be 
P^\alpha(T) = |\langle \psi_g| \psi^\alpha_g(\tau = 1)\rangle|^2,
\ee
for $\alpha \in \{F, A, M\}$. In the case of degenerate final ground states, we redefine the success probability as the sum over individual success probabilities with equal weights.

We study systems of $N$ spins where $6 \leq N \leq 17$ and choose a fixed annealing time of $T = 100$ for all system sizes. For each system size we generate 10000 random instances of the problem and for each instance calculate the success probabilities and the instantaneous energy spectra as functions of time according to our four annealing schedules,  Eqs.~(\ref{eq:HX}, \ref{eq:HXX}). 

To better compare the performance of different Hamiltonians, we define two additional quantities. The first is the {\it success probability enhancement ratio}, defined for each type of Hamiltonian with coupled drivers $H^\alpha$ as the percentage of instances for which $H^\alpha$ provides the best improvement over $H^0$, i.e. it performs better than $H^0$, as well as the other two Hamiltonians with coupled drivers. If we denote the number of such instances with $L^\alpha$ and the total number of instances with $L = 10000$ then the enhancement ratio is simply defined as,
\be
R_{en}^{\alpha} = \frac{L^\alpha }{L}. 
\label{eq:Ren}
\ee
For each of the instances identified in $R_{en}^\alpha$, we then define the {\it success probability enhancement}, which measures the actual enhancement that results from applying $H^\alpha$, i.e. 
\be 
P_{en}^{\alpha} = \frac{P^\alpha}{P^0}, 
\label{eq:Pen}
\ee
Note that we always have $P_{en}^{\alpha} > 1$.

\section{Numerical Results}
\label{sec:results}

\subsection{Success probability Enhancement}

\begin{figure*}
\begin{center}
\includegraphics[width = 2\columnwidth]{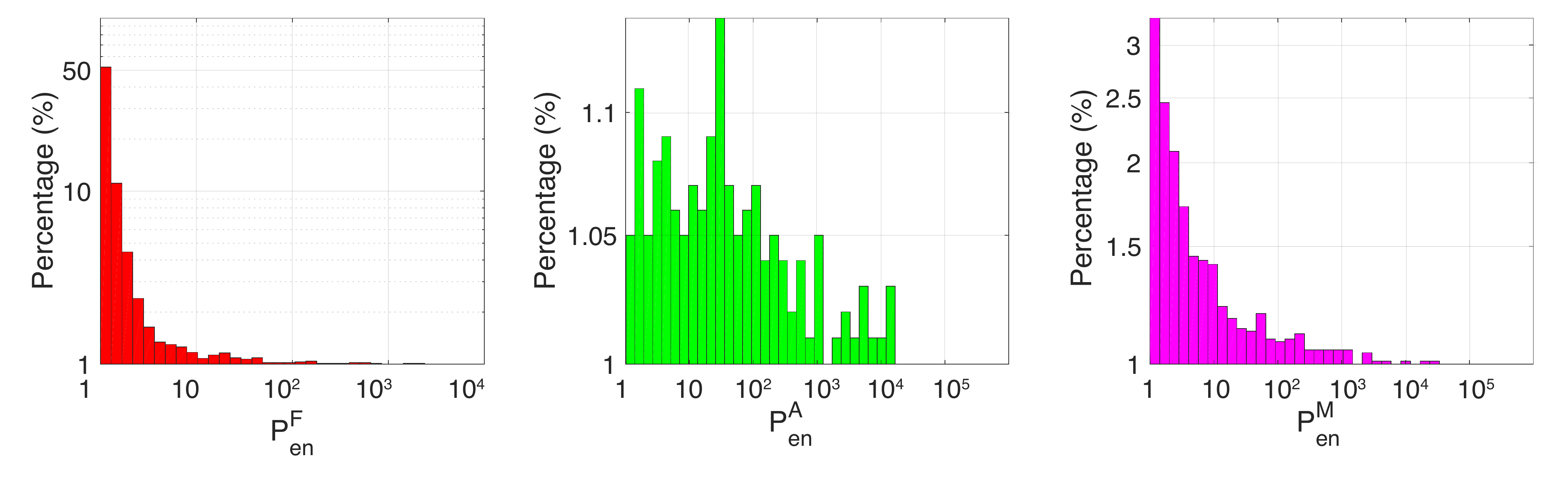} 
\includegraphics[width = 2\columnwidth]{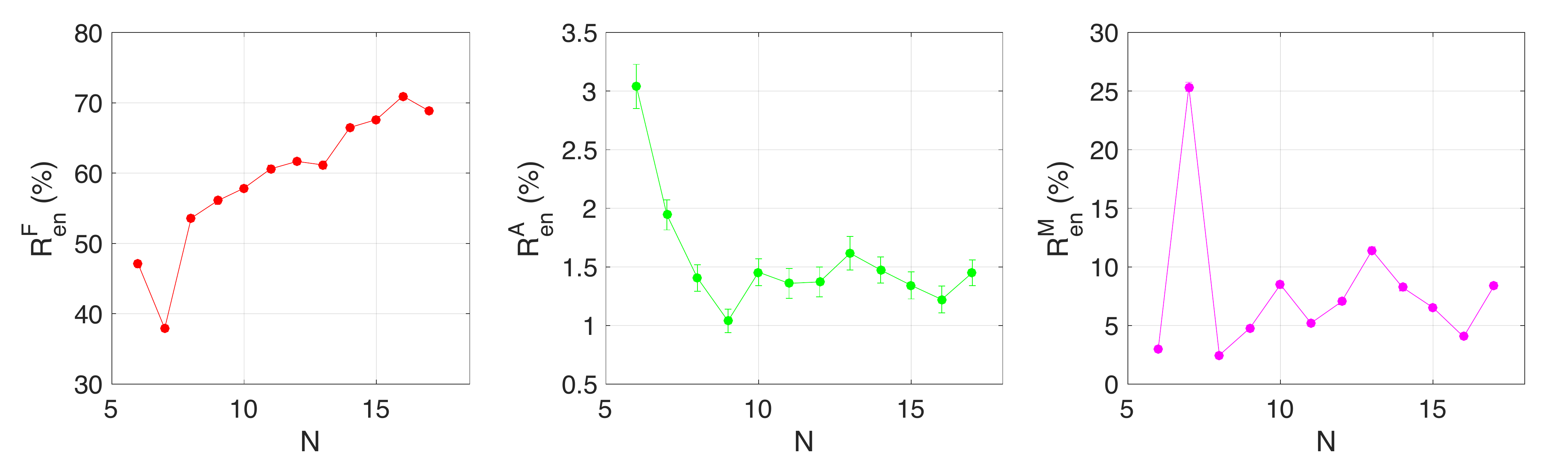} 
\includegraphics[width = 2\columnwidth]{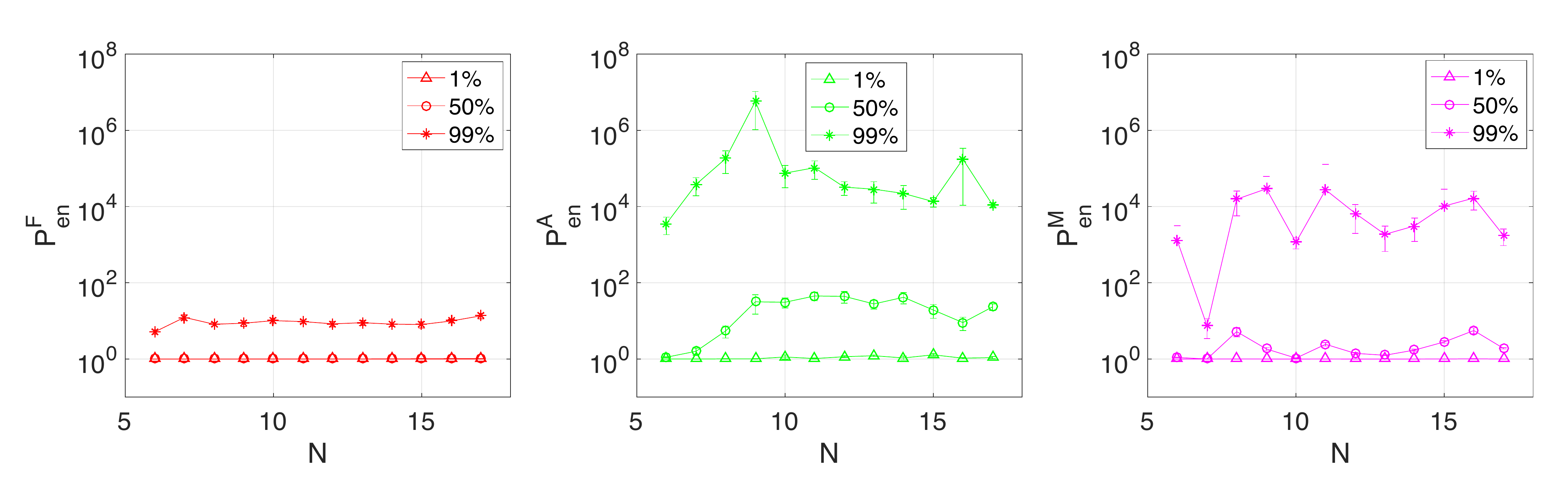} 
\caption{({\bf Top Panel}) The distribution of success probability enhancement, $P_{en}^\alpha$ for $\alpha \in \{F, A, M\}$, resulting from the three Hamiltonians with coupled drivers $H^\alpha$in a system of $N = 17$ spins. 
({\bf Middle Panel}) Success Probability Enhancement Ratio, $R_{en}^\alpha$, as a function of system size for the three types of coupled driver Hamiltonians. 
({\bf Bottom Panel}) The  $1^{st}$, $50^{th}$ and $99^{th}$ percentile values of  $P_{en}^\alpha$ for each  Hamiltonian with coupled drivers as a function of system size. 
}
\label{fig:Pen}
\end{center}
\end{figure*}

\begin{figure*}
\begin{center}
\includegraphics[width=2\columnwidth]{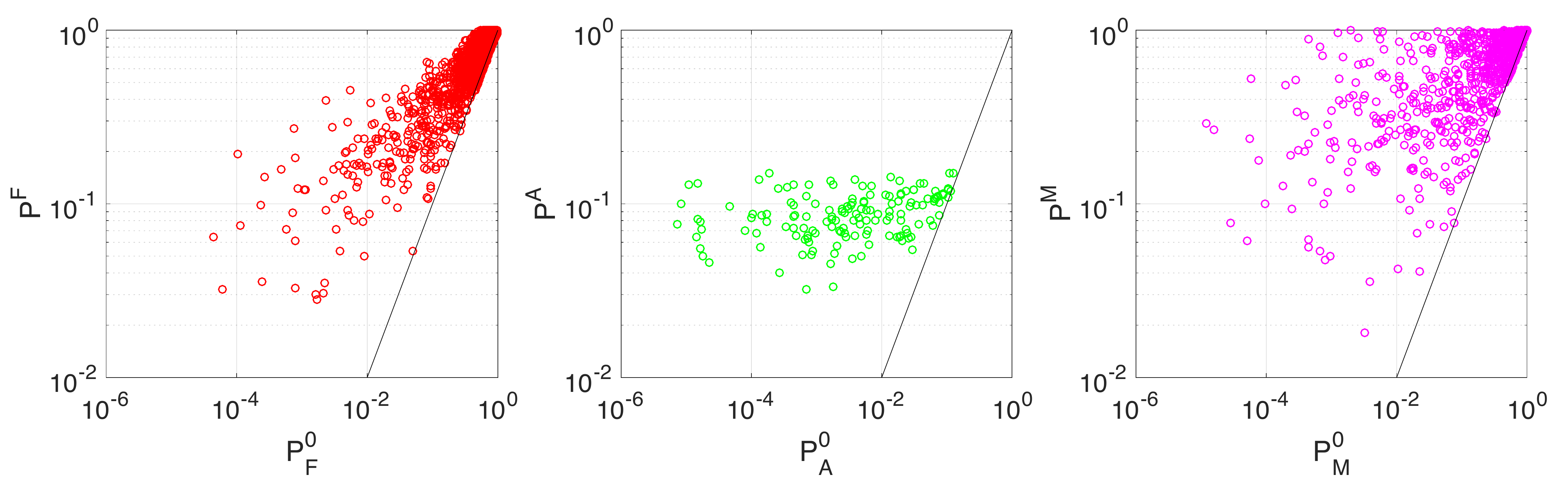} 
\includegraphics[width = 2\columnwidth]{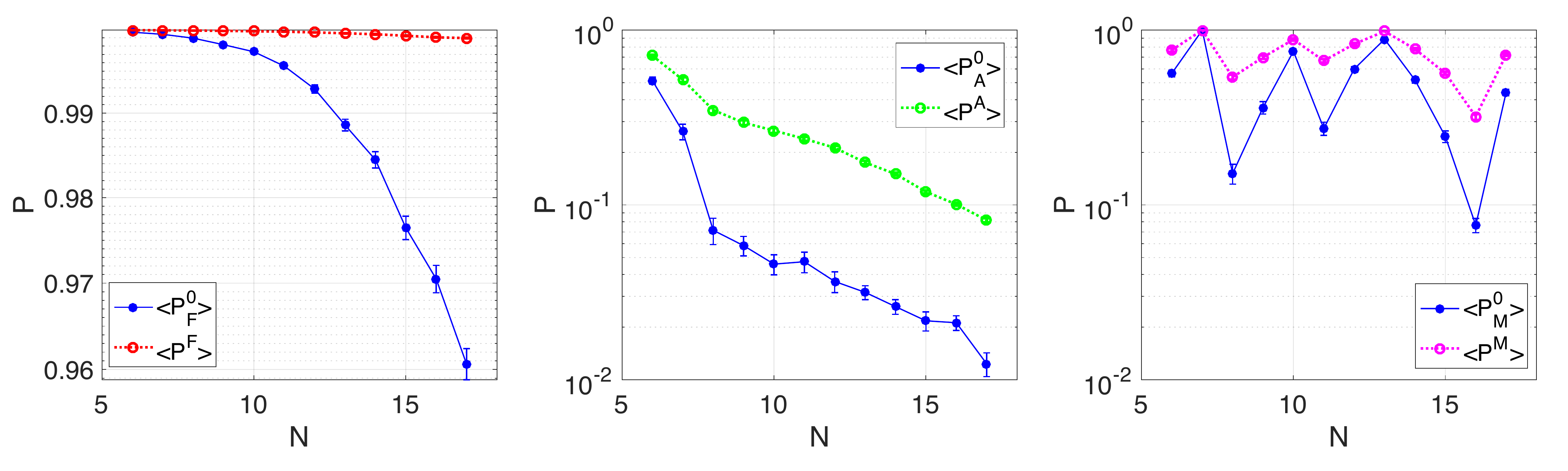} 
\caption{Probability distribution of those instances that show improvement in the presence of Hamiltonians with coupled drivers, $H^\alpha$ with $\alpha \in \{F, A, M\}$ compared to the probabilities resulting from the single spin flip Hamiltonian, $H^0$. 
({\bf Top Panel}) Scatter plots of $P^0_\alpha$, the probabilities resulting from $H^0$ of the instances that show the best improvement once $H^\alpha$ is used vs. $P^\alpha$, the probabilities of those same instances now resulting from $H^\alpha$. 
({\bf Bottom Panel}) The median values of the success probability distributions of the affected instances, resulting from $H^0$  (shown in blue) and those resulting from $H^\alpha$ (red, green, and magenta), as functions of system size. 
}
\label{fig:PAI}
\end{center}
\end{figure*}

\begin{figure*}
\begin{center}
\includegraphics[width=2\columnwidth]{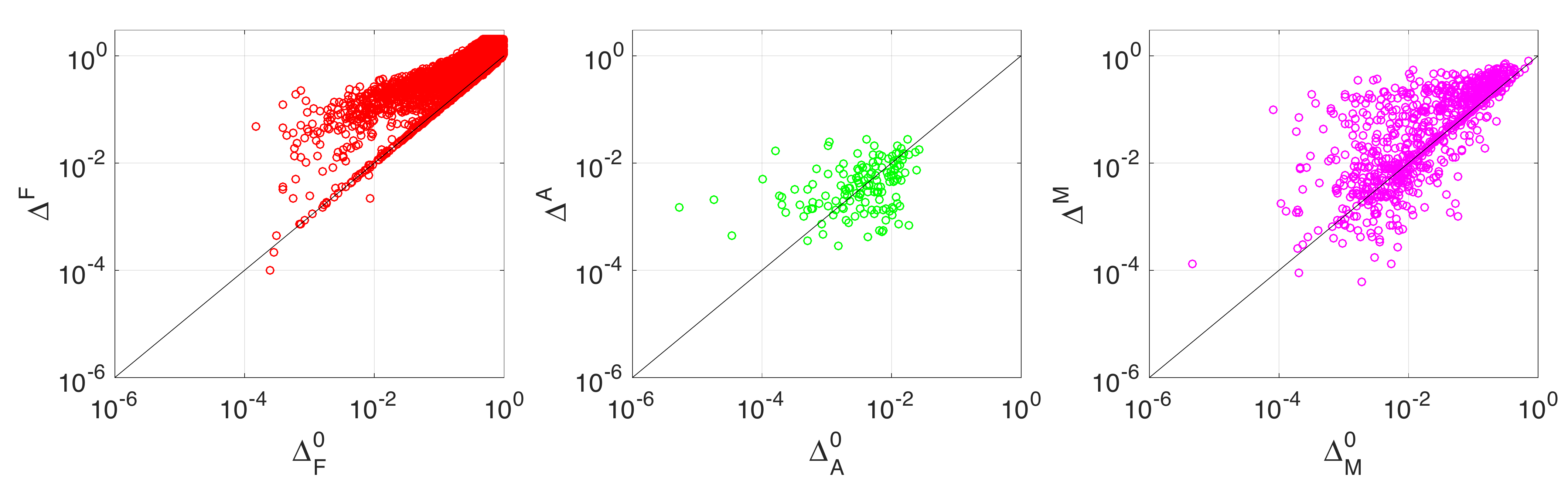} 
\includegraphics[width = 2\columnwidth]{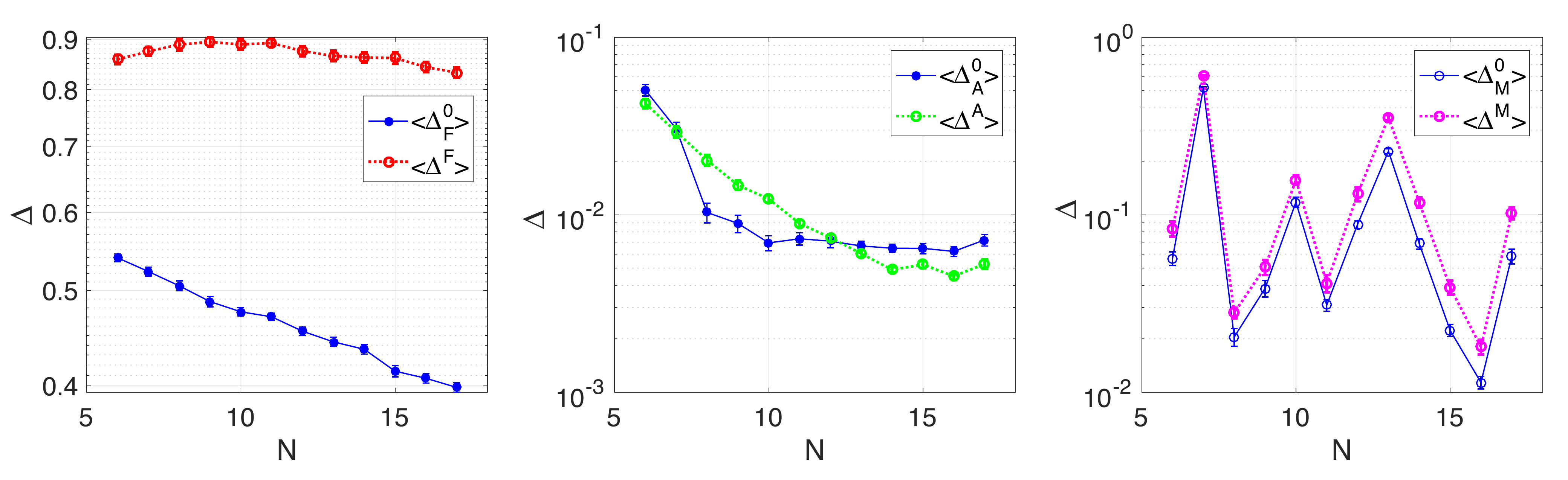} 
\caption{Distribution of minimum gap for the affected instances that show the best improvement in success probability using Hamiltonians with coupled drivers $H^\alpha$ with $\alpha \in \{F, A, M\}$. 
({\bf Top Panel}) Scatter plots of the initial minimum gaps of the affected instances, $\Delta^0_\alpha$, resulting from $H^0$, vs. final minim gaps $\Delta^\alpha$, resulting from $H^\alpha$ for each Hamiltonian with coupled drivers. 
({\bf Bottom Panel}) The median values of the minimum gap distributions of the affected instances, resulting from $H^0$  (shown in blue) and the coupled driver Hamiltonians (red, green, and magenta), as a function of system size. 
}
\label{fig:MinGapAI}
\end{center}
\end{figure*}

\begin{figure*}
\begin{center}
\includegraphics[width = 2\columnwidth]{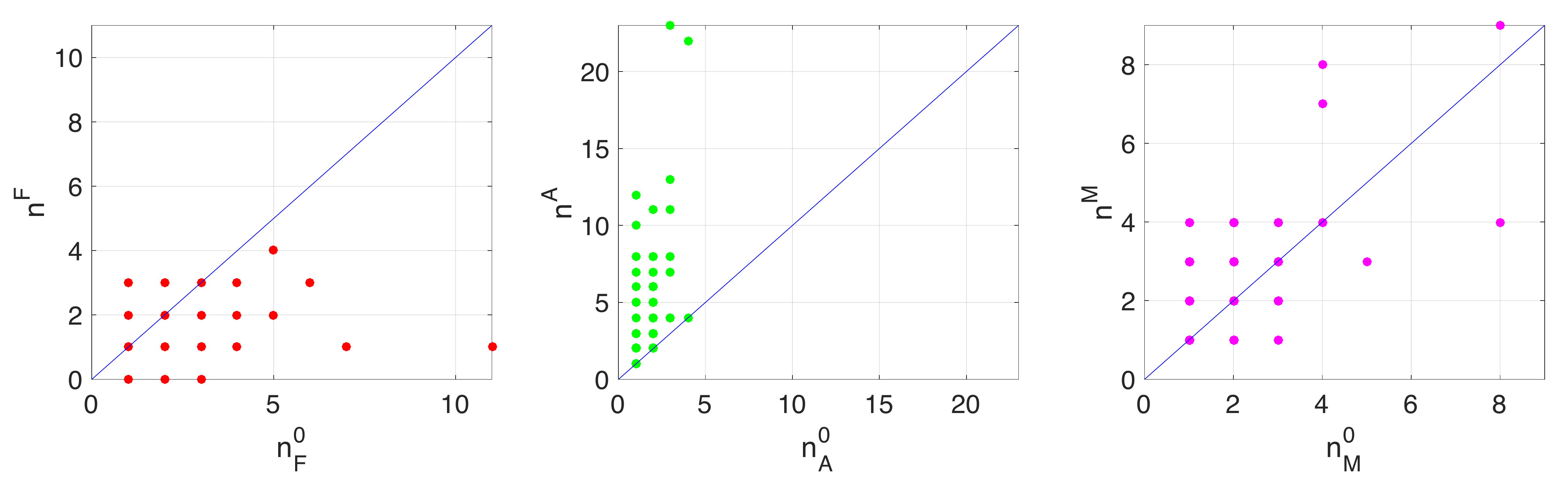} 
\includegraphics[width = 2\columnwidth]{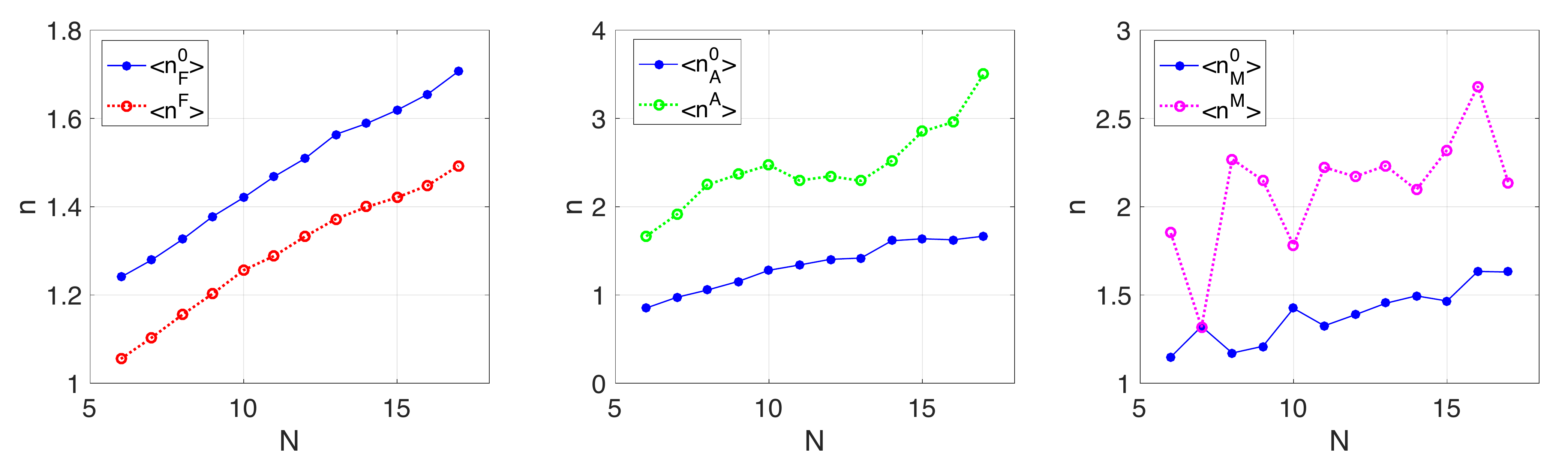} 
\caption{
Distribution of the number of anticrossings for the affected instances that show the best improvement in success probability using Hamiltonians with coupled drivers $H^\alpha$ with $\alpha \in \{F, A, M\}$. 
({\bf Top Panel}) Scatter plots of the initial number of anticrossings of the affected instances, $n^0_\alpha$, resulting from $H^0$, vs. their final number $n^\alpha$, resulting from $H^\alpha$ for each Hamiltonian with coupled drivers. 
({\bf Bottom Panel}) The mean number of anticrossings of the affected instances, resulting from $H^0$  (shown in blue) and $H^{F, A, M}$ (shown in red, green, and magenta), as functions of system size. 
}
\label{fig:nlcAI}
\end{center}
\end{figure*}

We start our analysis by determining the success probability enhancement ratios, $R_{en}^\alpha$, and the corresponding enhancements, $P_{en}^\alpha$, for each Hamiltonian with coupled drivers, $H^\alpha$.
For a system of $N = 17$ spins we find that the stoquastic Hamiltonian has a large enhancement ratio of $R_{en}^F \simeq 68.8\%$ while the nonstoquastic Hamiltonians produce much smaller ratios of $R_{en}^A \simeq 1.5\%$ and $R_{en}^M \simeq 8.4\%$. 

The top panels of Fig.~\ref{fig:Pen} show the distributions of the corresponding enhancements, $P_{en}^\alpha$. We see that  for the stoquastic Hamiltonian the distribution is very uneven with a sharp peak near the unity and a very modest $99^{th}$ percentile value of order $O(10)$. In contrast, for the nonstoquastic Hamiltonian $H^A$ we see that the distribution is substantially more spread-out with the $99^{th}$ percentile value of the enhancement being of the order $O(10^4)$. The enhancement distribution for the other nonstoquastic Hamiltonian $H^M$ is peaked near unity but it also has a fat tail with the $99^{th}$ percentile value of  order $O(10^3)$.

We then studied the dependence of $R_{en}^\alpha$ on system size. The results, shown in the second row of Fig.~\ref{fig:Pen}, indicate that for the stoquastic Hamiltonian $R_{en}^F$ remains large and it even grows with the system size from $R_{en}^F \simeq 47\%$ for $N = 6$ to $R_{en}^F \simeq 69\%$ for $N = 17$. For the nonstoquastic Hamiltonian with uniform antiferromagnetic couplings we see that $R_{en}^A$ initially decreases and then saturates around $R_{en}^A \simeq 1.5\%$  while for the other nonstoquastic Hamiltonian with mixed-sign couplings $R_{en}^M$ fluctuates around a mean value of $R_{en}^M \simeq 8\%$. We attribute the large fluctuations in $R_{en}^M$ to the random nature of sign assignments in the coupled driver term, $H^M_I$, and the fact that for each system size a different ratio of plus to minus signs is assigned~\cite{fn:random}. As will be shown in the following sections, these fluctuations remain manifest in other quantities that result from $H^M$ as well.

We next study the change in the distribution of success probability enhancement as a function of system size by plotting the $1^{st}$, $50^{th}$ and $99^{th}$ percentile values of  $P_{en}^\alpha$ as functions of $N$. These plots (shown in the bottom panels of Fig.~\ref{fig:Pen}) indicate that the distributions of success probability enhancement remain fairly constant as the system size grows. For the stoquastic Hamiltonian, $H^F$, we see that the distribution remains peaked near minimal enhancement, as indicated by the $1^{st}$ and $50^{th}$ percentile values lying close to each other near unity, and a modest $99^{th}$ percentile with average value of $\langle P_{en}^{F \; 99\%} \rangle \simeq 10$ for all system sizes. For the nonstoquastic Hamiltonians, $H^A$ and $H^M$ we see that the distributions remain spread-out across different system sizes, with the $99^{th}$ percentile values of the enhancements persistently fluctuating around much larger average values of $\langle P_{en}^{A \; 99\%} \rangle \simeq O(10^5)$ and $\langle P_{en}^{M \; 99\%} \rangle \simeq O(10^4)$.

As in the case of $N = 17$, for all system sizes we see a clear difference between the improvement due to stoquastic and nonstoquastic Hamiltonians: the stoquastic Hamiltonian improves a large fraction of instances and this fraction grows with the system size, but the actual enhancement due to this Hamiltonian is modest. In contrast, the nonstoquastic Hamiltonians affect much smaller fractions of instances, which remain fairly constant as the system size grows, but the actual enhancements can be very large. This trade-off between the enhancement ratio and the corresponding enhancement in success probability occurs in other Hamiltonians that we have studied as well~\cite{fn:other}.

\subsection{Probability Distribution of Affected Instances}

We next take a closer look at the specific instances that show the best success probability under each type of Hamiltonian with coupled drivers, $H^\alpha$, and call them \emph{affected instances}. Note that, given this definition, there are exactly three non-overlapping~\cite{fn:AI} sets of affected instances, one for each $\alpha \in \{F, A, M\}$. The goal here is to determine the common properties of each set of affected instances and to classify them based on how hard they are for $H^0$ and the best possible improvement that can be obtained from the corresponding $H^\alpha$. 

Thus, we are interested in the \emph{initial} probabilities of the affected instances, resulting from $H^0$, and the \emph{final} probabilities of the same instances resulting from $H^\alpha$. To make referencing easier, for each set of affected instances, we denote the initial probabilities with $P_\alpha^0$, where the additional subscript $\alpha$ refers to the Hamiltonian with coupled drivers for which the affected instances show the best final success probabilities, $P^\alpha$. 

The panels in the top row of Fig.~\ref{fig:PAI} show scatter plots of the initial and final probabilities of the affected instances, for a system of $N = 17$ spins. Here the vertical axes correspond to the final probabilities $P^\alpha$ of the affected instances resulting from $H^\alpha$ and the horizontal axes represent the initial probabilities $P^0_\alpha$ of those same instances resulting from $H^0$. We see that the stoquastic Hamiltonian $H^F$ affects a large range of instances, but mainly those with higher initial probabilities as is evident by the concentration of the instances near $P^0_F \simeq 1$. 

For the nonstoquastic Hamiltonian $H^A$ we see that the range of initial probabilities is very small and is limited to very hard problems. As a result of applying $H^A$, the lower bound improves substantially while the upper bound shows little improvement. Finally for $H^M$ we see that problems with a large range of initial probabilities can be improved, including both easy and hard problems, and the resulting probabilities also cover a large range.

To determine the finite-size effects in the initial and final probabilities, we plot the median values of $P^0_\alpha$ and $P^\alpha$ for various system sizes. These plots, depicted in the bottom panels of Fig.~\ref{fig:PAI}, show that the problems get harder as the system size grows, as is expected. Furthermore, the improvement in success probability continues to be significant in the case of the nonstoquastic Hamiltonians $H^A$ and $H^M$ while it remains marginal for the stoquastic Hamiltonian $H^F$. 

So far we have seen that the stoquastic Hamiltonian tends to provide small improvements to easier problems, whereas nonstoquastic Hamiltonians mainly provide larger improvements to harder problems. To gain a better understanding of the mechanisms behind the performance of each type of Hamiltonian, we next study the instantaneous energy spectrum of the system, which we  calculate using exact diagonalization.

\subsection{Relation to the Size of Minimum Gaps}

We first consider the distribution of minimum gaps for the affected instances of each Hamiltonian. For each set, we are interested in the distributions of initial minimum gaps, resulting from $H^0$, and final minimum gaps, resulting from $H^\alpha$. We use a similar notation to the case of success probabilities and for each set of affected instances denote the initial minimum gaps with $\Delta^0_\alpha$ and the final values with $\Delta^\alpha$. Scatter plots of these values for a system of $N = 17$ spins are shown in the top panels of Fig.~\ref{fig:MinGapAI}. Similar to the plots of probabilities, here too the vertical axes correspond to the final minimum gaps and the horizontal axes represent the initial minimum gaps of the affected instances.

For the stoquastic Hamiltonian, we see that the affected instances cover a large range of initial gaps and the addition of $H_I^F$ increases the gap for the great majority of the affected instances.  For the nonstoquastic Hamiltonian $H^A$, we see that the range of the initial gaps is smaller and turning on $H_I^A$ mainly improves the lower bound. Note that the final gaps $\Delta^A$ can increase but also for about half of the instances $\Delta^A \leq \Delta^0_A$. The effect of $H^M$ on the minimum gaps of its respective affected instances seems to be similar to both $H^F$ and $H^A$: the initial gaps cover a large range of values, and while the application of $H^M$ results in a modest increase in the size of the gaps for the majority of instances, still for a significant number of them the gap decreases or remains unchanged. 

Plots of the median values of minimum gaps, resulting from $H^0$ and $H^\alpha$, as functions of system size are depicted in the bottom panels of Fig.~\ref{fig:MinGapAI}. These plots show that the trends observed in the case of $N = 17$ remain persistent as the system size grows, i.e. the stoquastic Hamiltonian almost always increases the minimum gaps of its affected instances by a significant amount, while for nonstoquastic Hamiltonians this is not the case. 

In the case of the stoquastic Hamiltonian a general increase in the size of the gap during the earlier stages of annealing is expected and can be explained using a mean-field model, where the ferromagnetic intermediate term effectively increases the strength of the transverse-field and thus also the overall gap. It is plausible that the same effect is also responsible for the observed increase in the size of the minimum gaps at the transition points for the instances that we have studied.

The increase in the size of the gaps provides a straightforward explanation for the observed improvement in the final success probabilities, as it reduces the likelihood of the system transitioning away from its ground state during the annealing process~\cite{LZ}. This seems to be the dominant mechanism by which the stoquastic Hamiltonian $H^F$ improves the success probability of the majority of its affected instances.

Note that in general, for the instances with extremely small initial gaps, one can expect that any perturbation to the annealing Hamiltonian, stoquastic or nonstoquastic, has a high chance of increasing the final gaps. The case for nonstoquastic Hamiltonians, however, remains enigmatic since for a significant number of the affected instances the gap does not increase. In the next section we study the energy spectrum more closely and shine some light on this puzzle.

\subsection{Relation to the Number of Anti-Crossings}

The final quantity that we consider is the number of anticrossings between the ground state and the first excited state energies during the evolution of the system. We study this quantity for the affected instances for each Hamiltonian with coupled drivers and use the notation $n^0_\alpha$ and $n^\alpha$ for the initial and final numbers of anticrossings resulting from $H^0$ and $H^\alpha$, respectively~\cite{fn:nlc}. 

The top panels of Fig.~\ref{fig:nlcAI} show scatter plots of these quantities for a system of $N = 17$ spins. For the example shown we see that the stoquastic Hamiltonian mainly reduces the number of anticrossings while the nonstoquastic Hamiltonians increase them. This is particularly evident in the case of $H^A$. Plots of the average numbers of anticrossings as functions of system size, shown in the bottom panel of Fig.~\ref{fig:nlcAI}, further confirm these observations for all systems sizes: for stoquastic Hamiltonians the average number of anticrossings slightly decreases, while for nonstoquastic Hamiltonians this quantity clearly increases. 

The increase in the number of anticrossings in the case of nonstoquastic Hamiltonians can be explained by noting that the presence of long-range anti-ferromagnetic couplings in these systems increases the level of frustration, thus modifying the corresponding instantaneous energy spectra with the addition of more anticrossings with small gaps. One can speculate that the increase in the number of anticrossing can  have a beneficial effect on the hardest instances.  For these instances, the minimum gap is generally very small so the system is very likely to transition away from the ground state during its evolution.  Modifying the spectrum by adding extra anticrossings with comparably small gaps provides the system with further opportunities to transition back to the ground state, thereby correcting the earlier errors and improving the final success probability. This phenomenon  is similar to the observation reported in Ref.~\onlinecite{crosson14}, where for some very hard instances of MAX 2SAT, it was found beneficial to start the annealing from the first excited state of the beginning Hamiltonian instead of the usual choice of the ground state. 

Note that, without prior knowledge of the spectrum, this mechanism can improve or worsen the final success probabilities on a random basis, hence it provides enhancement for only a small number of lucky instances that can take advantage of it. Nevertheless, since the initial success probabilities in these cases are often very small, the resulting improvement due to this process can be significant and this is consistent with our observations.

\section{Discussion and Conclusions}
\label{sec:discussion}

In this work we have provided a systematic analysis of the performance of quantum annealers with stoquastic and nonstoquastic Hamiltonians, in finding the ground state of long-range Ising spin glass problems. We first constructed two different nonstoquastic Hamiltonians by adding purely antiferromagnetic, and mixed ferromagnetic and antiferromagnetic driver terms of the form $\sigma^x\sigma^x$ to the annealing schedule. We then compared their performance against the performances of a stoquastic Hamiltonian with ferromagnetic couplings, as well as a pure transverse-field Hamiltonian. We observed that, for subsets of instances of our spin glass problem, both stoquastic and nonstoquastic Hamiltonians with coupled driver terms outperform the traditional transverse-field only quantum annealers, however, the resulting enhancements are qualitatively different for the two classes. 

For stoquastic Hamiltonian, $H^F$, we observed that the fraction of the affected instances is large and it increases as the system size grows. A closer look at the specific instances for which $H^F$ provides the best improvement reveals that the majority of such instances can be easily solved by $H^0$, and the addition of the extra coupling terms in $H^F$  provides only marginal improvement to the final success probabilities. An examination of the minimum gaps for these instances reveals that, for most, the initial gaps are large and they further increase once $H^F$ is applied. Finally we saw that the numbers of anticrossings between the first two energy levels decrease for most instances. The general decrease in the number of anticrossings, and the increase in the size of the gap, can be explained using a mean-field description of $H^F$, which provides a straightforward explanation for the enhanced performance of the stoquastic Hamiltonian. 

For nonstoquastic Hamiltonians, we saw that the fractions of affected instances are much smaller than the stoquastic case, and that they remain relatively constant as the system size varies. In this case we noticed that the majority of affected instances are hard for $H^0$ and that the addition of the extra coupling terms can provide significant improvements to the initial success probabilities. We also observed that in this case the average minimum gap does not change significantly, but the average number of anticrossings clearly increases. We argued that the increase in the number of anticrossings can, on a random basis, significantly improve the success probability of the hardest instances with tiny minimum gaps.  

This work is a starting point for a series of deeper investigators into the potential advantages of nonstoquastic Hamiltonians and the mechanisms responsible for their performance.  A promising future direction is to carry out a more detailed study of nonstoquastic Hamiltonians with mixed ferromagnetic and antiferromagnetic couplings. It would be interesting to understand whether optimized couplings can lead to further improvement in the performance of quantum annealers and to probe deeper into their inner workings.

Other generalizations of this work along various lines can also be foreseen. For example, it would be important to study the effect of optimizing both the annealing schedule and the annealing time using feedback from the performance of quantum annealers for various problem Hamiltonians. It would also be interesting to study the performance of other nonstoquastic Hamiltonians where the driving terms are of the form $\sigma^x\sigma^z$.  These terms can naturally emerge in certain qubit architectures, but their effect on the annealing process is not yet determined. Finally, one should go beyond the unitary dynamics of this work and consider the effects of interactions with the environment and couplings with a dissipative bath to assess their impact in a realistic setup.
 
\acknowledgments

Discussions and correspondences with S. Bravyi, P. Cappellaro, C. Chamon, E. Crosson, E. Farhi, A. Harrow, H. Katzgraber, K. Kechedzhi, J. Kerman, S. Knysh, H. Nishimori,
A. Polkovnikov, A. Sandvik, and I. Zintchenko at various stages of this work are gratefully acknowledged. We are also grateful to P. Love and S. Mandra for discussions and comments on an earlier draft of the manuscript. This work has been supported by the Swiss National Science Foundation through the National Competence Center in Research QSIT, by IARPA via MIT Lincoln Laboratory Air Force 538 Contract No. FA8721-05-C-0002, and by a grant from the Swiss National Supercomputing Centre (CSCS) Project ID S686. The views and conclusions contained herein are those of the authors and should not be interpreted as necessarily representing the official policies or endorsements, either expressed or implied, of ODNI, IARPA, or the US Government. The US Government is authorized to reproduce and distribute reprints for Governmental purpose notwithstanding any copyright annotation thereon.


\begin{references}
\bibitem{fn:zeroT} Here we focus on situations where the temperature is well below the minimum gap, hence effectively zero.
\bibitem{fn:qmcmarkov} This does not mean, however, that the mixing of the QMC Markov chain is fast and efficient or that the ground state can be found quickly.
\bibitem{fn:SK} In the original Sherrington-Kirpatrick model $h_i$ = 0 and $J_{ij}$ are chosen from a distribution with variance proportional to $1/N$ (see Ref.~\cite{SK}).
\bibitem{fn:scaling}  If we normalize the Hamiltonians then $\lambda = 1/N$.
\bibitem{fn:random} Here we apply the same random sign assignment to all instances within a system size, hence the error bars are small while the fluctuations between different system sizes are large. To generate error bars comparable to the fluctuations between system sizes one has to instead assign random sign configurations to each instance.
\bibitem{fn:other} In addition to the random configuration of plus and minus signs in $H_I^M$ (Eq.~\ref{eq:HiXXS}) reported here, we have also studied two other intermediate Hamiltonians with different combinations of plus and minus signs: one where $r_{ij} = (-1)^{(iN + j)}$ and the other where we correlate the signs with the signs of the couplings in the problem Hamiltonian i.e. $r_{ij} = \text{sgn}(J_{ij})$. The results are similar to $H_I^M$ but the latter performs slightly better.
\bibitem{fn:nlc} Here we calculate the number of anticrossings by counting the local minima in plots of the instantaneous energy gap as a function of time. We calculate the energy gap, using exact diagonalization, for 100 time steps so the accuracy of the number of anticrossings is limited to the size of these time steps.
\end{references}
\end{document}